\documentclass[]{aa}  

\usepackage{graphicx}
\usepackage{txfonts}
\usepackage[]{hyperref}
\usepackage{color}
\usepackage{amsmath}	% Advanced maths commands
\usepackage{amssymb}	% Extra maths symbols
\usepackage{multicol}        % Multi-column entries in tables
\usepackage{bm}		% Bold maths symbols, including upright Greek
\usepackage{pdflscape}	% Landscape pages
\usepackage{multirow}
\usepackage[T1]{fontenc}
\usepackage{comment}
\usepackage{pifont}

\newcommand{\oneh}{1H~0707-495}

\newcommand{\xmm}{{\em XMM-Newton}}
\newcommand{\nus}{{\em NuSTAR}}

\newcommand{\expo}[2]{$ #1 \times 10^{#2}$}
\newcommand{\tento}[1]{$10^{#1}$}

\newcommand{\expom}[2]{ #1 \times 10^{#2}}

\newcommand{\msun}{M_{\odot}}

\newcommand{\ledd}{L_{\textrm{Edd}}}
\newcommand{\eedd}{\epsilon_{\textrm{Edd}}}

\newcommand{\lumcgs}{erg~s$^{-1}$}

\newcommand{\sqcm}{cm$^{-2}$}

\newcommand{\nh}{N_{\textrm{H}}}

\newcommand{\rg}{$R_{\textrm{G}}$}

\newcommand{\subrm}[1]{_{\textrm{#1}}}

\newcommand{\fek}{Fe~K~$\alpha$}

\newcommand{\compps}{{\sc compps}}
\newcommand{\nthcomp}{{\sc nthcomp}}

\newcommand{\kyn}{\textsc{kynbb}}

\newcommand{\monk}{\textsc{monk}}

\defcitealias{dovciak&done2016}{DD16}
\newcommand{\ded}{\citetalias{dovciak&done2016}}

\begin{document} 

\title{Estimating the size of X-ray lamppost coronae in active galactic nuclei}

\titlerunning{The size of X-ray lamppost coronae in AGNs}
\authorrunning{F. Ursini et al.}

\author{
	F. Ursini\inst{1},
		M. Dov\v{c}iak\inst{2},
		W. Zhang\inst{2},
		G. Matt\inst{1}, 
		P.-O. Petrucci\inst{3}
		\and
		C. Done\inst{4}
}

\institute{
	Dipartimento di Matematica e Fisica, Universit\`a degli Studi Roma Tre, via della Vasca Navale 84, 00146 Roma, Italy\\
	\email{francesco.ursini@uniroma3.it}
	\and
	Astronomical Institute, Academy of Sciences of the Czech Republic, Bo\v{c}n\'i II 1401, CZ-14100 Prague, Czech Republic
	\and
	Univ. Grenoble Alpes, CNRS, IPAG, 38000 Grenoble, France
	\and 
	Centre for Extragalactic Astronomy, Department of Physics, University of Durham, South Road, Durham, DH1 3LE, UK
}

\date{Received ...; accepted ...}

% \abstract{}{}{}{}{} 
% 5 {} token are mandatory

\abstract
% context heading (optional)
% {} leave it empty if necessary  
{}
% aims heading (mandatory)
{We report estimates of the X-ray coronal size of active galactic nuclei in the lamppost geometry. In this commonly adopted scenario, the corona is assumed for simplicity to be a point-like X-ray source located on the axis of the accretion disc. However, the corona must intercept a number of optical/UV seed photons from the disc consistent with the observed X-ray flux, which constrains its size.  }
% methods heading (mandatory)
{We employ a relativistic ray-tracing code, originally developed by \cite{dovciak&done2016}, that calculates the size of a Comptonizing lamppost corona illuminated by a standard thin disc. We assume that the disc extends down to the innermost stable circular orbit of a non-spinning or a maximally spinning black hole. We apply this method to a sample of 20 Seyfert 1 galaxies, using simultaneous optical/UV and X-ray archival data from \xmm.}
% results heading (mandatory)
{At least for the sources accreting below the Eddington limit, we find that a Comptonizing lamppost corona can generally exist, but with constraints on its size and height above the event horizon of the black hole depending on the spin.
For a maximally spinning black hole, a solution can almost always be found at any height, while for a non-spinning black hole the height must generally be higher than 5 gravitational radii.
This is because, for a given luminosity, a higher spin implies more seed photons illuminating the corona due to a larger and hotter inner disc area. The maximal spin solution is favored, as it predicts an X-ray photon index in better agreement with the observations.  }
% conclusions heading (optional), leave it empty if necessary 
{}


\maketitle
%
%-------------------------------------------------------------------

\section{Introduction}
\label{sec:intro}
The X-ray emission of active galactic nuclei (AGNs) is thought to be produced via thermal Comptonization of optical/UV photons, emitted by the accretion disc, in a hot corona \cite[e.g.][]{haardt&maraschi1991}.
A tight correlation is found between the X-ray and UV luminosity in quasars, indicating the existence of an universal coupling between the disc and the corona \citep{lusso&risaliti2016,lusso&risaliti2017}. The nature of this coupling is not fully understood \citep[e.g.][]{arcodia2019}, however a possible physical explanation could be the heating of the corona via reconnection of magnetic fields above the disc  \citep[and references therein]{lusso&risaliti2017}.

The geometry of the hot X-ray corona (i.e. its size and location) is poorly known and a matter of debate. Arguably the best constraints can be obtained from the analysis of microlensing variability, although this method is currently limited to a few strongly lensed quasars \citep[e.g.][]{chartas2002,chartas2009,chartas2016}. Another promising method relies on the X-ray spectral-timing properties of nearby bright Seyfert galaxies, especially from the detection of X-ray reverberation lags \citep{fabian2009,cackett2014,kara2016,demarco&ponti2019}.
Such observational constraints are generally consistent with a compact 
X-ray source, located within a few gravitational radii of the black hole \cite[e.g.][]{reis&miller2013}.
In a few bright Seyfert galaxies, the spectral modeling of the 
optical-to-X-ray continuum also yields similar estimates of the coronal size \cite[e.g.][]{pop2013mrk509,done2013,porquet2019}. X-ray polarimetry is also a promising technique to constrain the geometry of the corona, which influences the polarization signal \cite[]{moca1}. High sensitivity polarimetric observations will become possible in the near future thanks to the Imaging X-ray Polarimetry Explorer \cite[\textit{IXPE};][]{ixpe}.

The lamppost geometry is a configuration often used to describe the disc-corona system, where
the corona is assumed to be a point-like 
X-ray source located on the symmetry axis of the disc \cite[e.g.][]{matt1991,MM1996,light-bending}.
This could be physically realized by collisions and shocks within 
an ejection flow or
a failed jet \cite[]{henri1991,hp1997,ghm2004}.
This geometry has been assumed in detailed models for the calculation of reflection spectra \cite[e.g.][]{rellinelp,relxill,NZ2016}
and tested in a number of sources showing reflection-dominated spectra, such as the narrow-line Seyfert 1 galaxies 1H~0707-495 \cite[]{zoghbi2010,fabian2012_1h,dauser2012,szanecki2020} and IRAS 13224-3809 \cite[]{ponti2010,chiang2015}. Interestingly, the observed properties of these objects are consistent with the presence of very compact X-ray sources, located close to the black hole.
The most extreme spectrum seen from \oneh\ would require a source lying within 1 gravitational radius of the event horizon of a rapidly spinning black hole (\citealt{fabian2012_1h}; see also \citealt{szanecki2020}). 

A small sized corona can be also radiatively compact, meaning a large ratio of the luminosity to the radius \cite[]{guilbert1983}. This in turns implies a large optical depth for pair production. Observational constraints on the coronal parameters so far obtained
suggest that pair production and annihilation may 
act as an effective thermostat controlling the coronal temperature \cite[]{fabian2015coronae,fabian2017}.

However, a Comptonizing corona 
must intercept a number of seed photons (per unit time) sufficient to explain the observed X-ray flux. This, for a given seed photon flux, implies a constraint on the solid angle subtended by the corona as seen from the disc.
\citet[][DD16 hereafter]{dovciak&done2016} developed a method to constrain the coronal size from the observed optical/UV and X-ray fluxes. \ded\ developed a relativistic ray-tracing code that 
estimates the radius of a spherical X-ray source, located on the symmetry axis of a standard accretion disc, that Comptonizes the soft disc photons. 
The aim of the present work is constraining the size of a Comptonizing lamppost corona 
in a statistically significant sample of AGNs using the \ded\ code. 

The paper is structured as follows. We present the sample in Sect. \ref{sec:sample}. In Sect. \ref{sec:main} we discuss the application of the \ded\ code and the main results. We summarize our conclusions in Sect. \ref{sec:conclusions}.

\section{The sample}\label{sec:sample}
In order to test the lamppost geometry with the \ded\ code, we need good constraints on both the disc luminosity and on the X-ray flux and spectral shape. We thus focus on bright, unobscured sources with simultaneous, high-quality data in the optical/UV and X-ray bands. The simultaneity prevents spurious effects due to inter-band variability.
\xmm\ is the best instrument to perform this kind of analysis, because it provides both high-quality X-ray spectra with the EPIC-pn
camera \cite[]{xmm_pn} and optical/UV data with the optical monitor \cite[OM;][]{xmm_om}.
We use the sample of Seyfert 1 galaxies discussed by \cite{pics}\footnote{\cite{pics} performed spectral analyses to test a ``two-corona'' model. In this scenario, the primary X-ray emission is produced in a hot corona, while the UV and soft X-ray excess below 1-2 keV are produced via Comptonization in a warm ($kT\sim$ 0.5 keV) corona above a nearly passive disc \cite[see also][]{rozanska2015,pop2020}. Although this model is different from the standard disc/hot corona assumed here, 
	the sample is perfectly suited for our purpose because the selection criteria are the same.}. This sample includes 22 objects observed by \xmm, with public data as of April 16, 2014, cross-correlated with the AGNs and quasars catalog of \cite{VV2010}. The sources are further selected using the criteria of the CAIXA catalog \cite[]{caixa1}, namely they are radio-quiet and unobscured (column density $\nh < \expom{2}{22}$ \sqcm). Moreover, each source is detected with at least four OM filters, to well constrain the optical/UV emission. Since the black hole mass is a required parameter in our case, our sample includes only the 20 sources for which a measurement of the black hole mass is available \cite[]{caixa1}, with a total of 96 observations. The basic properties of the sources are reported in Table \ref{tab:sample}. 

\begin{table*}
	\begin{center}
		\caption{Basic data of the Seyfert 1 galaxies in our sample \citep[see also][]{pics}. In the fourth column we report the number of pointings with \xmm\ as of April 16, 2014. $M\subrm{BH}$ is the black hole mass in solar masses. \label{tab:sample}}
		\begin{tabular}{lccccccc}
			\hline \hline
			Name  & Redshift & $\log M_{\textrm{BH}}$ & obs. & $\Gamma$ & $L\subrm{UV}$ (5--7 eV) & $L\subrm{X}$ (2--10 keV)&  $L\subrm{X}/L\subrm{Edd}$\\
			&&&&&(\tento{43} \lumcgs)&(\tento{43} \lumcgs)&(\tento{-3})\\
			\hline
			1H 0419-577  & 0.1040 & 8.58  &  8 & 1.50--1.85 & 38.0--62.8 & 21.9--42.2&4.6--8.8\\
ESO 198-G24 & 0.0455 & 8.48 & 3 &1.72--1.84&1.5--2.9&4.6--6.2&1.2--1.7\\
HE 1029-1401&0.0858&8.73&2&1.80--1.94&75.5--77.5&20.2--33.2&3.0--4.9\\
IRASF 12397+3333&0.0435&6.66&2&2.05--2.3&0.2--0.3&1.9--2.2&32.6--38.6\\
MRK 279&0.0304&7.54&3&1.78--1.85&3.2--3.4&5.1--5.6&11.7--12.7\\
MRK 335&0.0257&7.15&3&1.52--1.96&2.9--3.8&0.5--0.7&2.9--4.2\\
MRK 509&0.0343&8.16&16&1.68--1.80&8.2--22.0&8.1--13.9&4.6--7.9\\
MRK 590&0.0263&7.68&2&1.67--1.77&0.2&0.6--1.0&1.1--1.6\\
MRK 883&0.0374&7.28&4&1.50--1.90&0.6--1.0&0.3--0.6&1.4--2.6\\
NGC 4593&0.0090&6.73&7&1.87--1.88&0.1--0.3&0.6&8.5-- 8.6\\
PG 0804+761&0.1000&8.24&2&1.50--2.00&108--119&13.8--23.8&6.4--11.1\\
PG 0844+349&0.0640&7.97&2&1.50--2.12&11.7--13.1&0.7--47.9&0.6--4.1\\
PG 1114+445&0.1438&8.59&12&1.51--1.96&19.4--24.9&8.4--19.0&1.7--3.9\\
PG 1116+215&0.1765&8.53&5&1.83--2.09&206--236&25.9--37.7&6.0--8.8\\
PG 1351+640&0.0882&7.66&3&1.50--2.01&18.3--23.3&0.7--1.2&1.3--2.1\\
PG 1402+261&0.1640&7.94&2&1.85--2.04&67.9--68.5&8.9--11.7&8.1--10.7\\
PG 1440+356&0.0790&7.47&4&2.21--2.39&18.7--22.5&2.6--4.8&6.8--12.8\\
Q0056-363&0.1641&8.95&3&1.50--2.07&60.9--108&14.8--19.4&1.3--1.7\\
RE 1034+396&0.0424&6.41&8&2.00--2.39&1.1--1.5&3.1--4.9&9.4--15.0\\
UGC 3973&0.0221&7.72&5&1.50--2.09&0.6--1.5&0.5--2.4&0.8--3.6\\
			\hline
		\end{tabular}
	\end{center}
\end{table*}

\section{Numerical procedure}\label{sec:main}

The \ded\ code assumes a standard accretion disc, producing a multicolor blackbody emission with a \cite{NT} temperature radial profile through the \textsc{kynbb} model \citep{kynbb}. The code calculates the photon flux received by the corona, taking into account all relativistic effects: the Doppler energy shift between the comoving disc frame and the corona, the gravitational energy shift, light bending and aberration \cite[see also][]{dovciak2014}.
It is assumed that a fraction $(1- e^{-\tau})$ of the incoming seed photons, where $\tau$ is the Thomson optical depth of the corona, gets scattered in the X-ray band.  
The Comptonized spectrum emitted by the corona is calculated with the \nthcomp\ model \citep{nthcomp1,nthcomp2}.
Part of the X-ray photons emitted by the corona illuminate the disc and their reprocessing slightly rises the temperature, however this has a negligible effect on the photon flux (\ded). 

From the observed UV and X-ray fluxes, the \ded\ code computes the seed photon flux at the corona $f_{BB}$ (found by integrating over the disc radius) and the X-ray photon flux $f_{X}$, in the local frame of the lamppost. 
The details of the calculations can be found in \ded. Here we just remind
the equation for the coronal radius $R_c$:
\begin{equation}\label{eq:size}
\pi (R_c/R\subrm{G})^2 = \frac{f_{X}}{f_{BB}} \frac{g_L}{1-e^{-\tau}}
\end{equation} 
where \rg $\equiv GM/c^2$ is the gravitational radius and 
$g_L$ is the relativistic energy shift between the corona and the observer\footnote{  
	The redshift factor $g_L$ is determined by the temporal component of the Kerr metric: $g_{tt}=-[1-2r/(r^2+a^2\cos\theta)]$ in Boyer-Lindquist coordinates and geometrized units.	
	Since the lamppost is at a height $h$ on the symmetry axis ($\theta=0$), we get $g_L  = \sqrt{|g_{tt}|} = \sqrt{1 - 2h/(h^2+a^2) }$.} \cite[see also][]{kynbb}.
The optical depth is computed from the X-ray photon index $\Gamma$ using the model \compps\ \cite[]{compps}, which includes a relativistic treatment for the electron temperature. 
For a given $\Gamma$, we use \compps\ to derive the optical depth of a spherical corona that produces a spectrum having that $\Gamma$, assuming a temperature of 100 keV (see \ded\ and Sect. \ref{subsec:par}). For example, photon indices $\Gamma = 1.5, 1.75$ and 2 correspond to optical depths $\tau = 1.8, 1.2$ and 0.85, respectively.

The size of a lamppost corona at a given height is thus constrained by the ratio of the observed photon fluxes. Equation (\ref{eq:size}) assumes the conservation of the number of photons. Pair production and annihilation are thus neglected, which is a reasonable approximation for low plasma temperatures ($< 511$ keV). However, pair processes could still play a role in limiting the maximum temperature that the corona can reach (see Sect. \ref{sec:conclusions}).  

It must be remarked that in \nthcomp\ the seed photons illuminate the corona isotropically, which is clearly an approximation given the disc-corona geometry.
This limitation has been discussed in detail by \cite{zhang2019}, who developed a more self-consistent Monte Carlo radiative transfer code to calculate Comptonized spectra in the Kerr spacetime (\monk). 
Compared with the \ded\ code, \cite{zhang2019} found slightly larger coronal sizes, depending on the optical depth $\tau$. This small discrepancy can be simply represented by a correction factor. We thus complemented the \ded\ computations with the correction provided by \monk, obtained as follows.
For a combination of black hole spin and corona height, we perform several Monk simulations with different values of $\tau$ assuming a corona radius of 1 \rg\  \cite[for spherical coronae, the energy spectrum is not sensitive to the size of the corona; see][]{zhang2019}. Given an observed value of the X-ray photon index $\Gamma_{\rm obs}$, we find out the simulated spectrum that has the same photon index. Then we derive the corona radius $R_{\rm DD16}$ with the \ded\ method using the input from the corresponding simulated spectrum. The correction factor is then 1 \rg/$R_{\rm DD16}$ and, in our case, we obtain values in the range 1.1--1.4 and slightly increasing with the height.

\subsection{Input physical parameters} \label{subsec:par}
\begin{figure}
   	\centering
   	\includegraphics[width=\hsize]{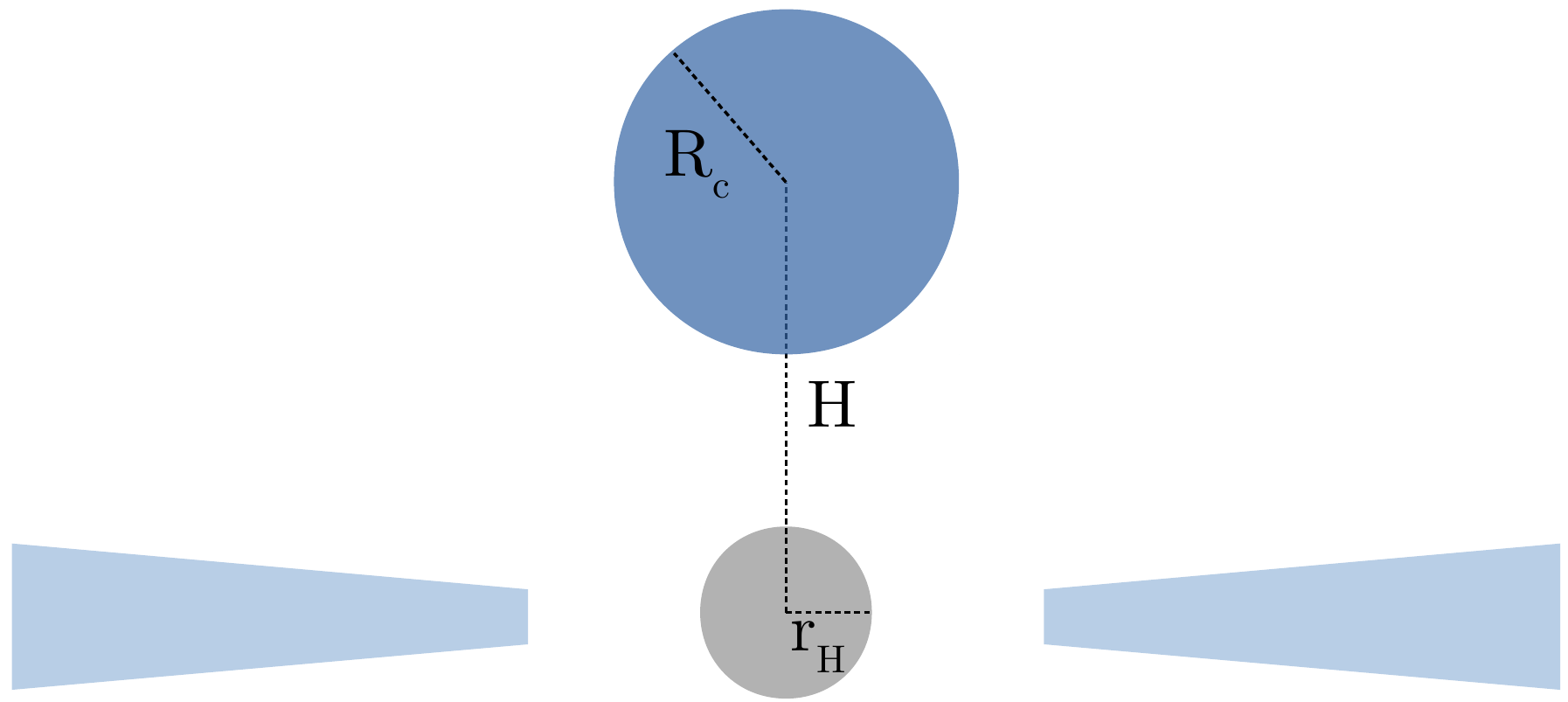}
   	\caption{Sketch of the lamppost configuration. The corona has a radius $R_c$ and is located at a height of $H-r_H$ above the event horizon.
   	}
   	\label{fig:a}
   \end{figure}

The crucial parameter determining the coronal size is the ratio between the number of Comptonized X-ray photons and the disc photons seen by the corona ($f_X/f_{BB}$ of eq. (\ref{eq:size})). As this ratio increases, the coronal size increases because the corona must intercept a larger number of disc photons. The photon number flux emitted by the disc and the corona can be easily inferred from the corresponding luminosity and spectral shape. For a Novikov-Thorne disc, the spectral shape is set by the inner disc radius, the accretion rate and the black hole mass. 
We assumed that the inner disc radius is equal to the innermost stable circular orbit (ISCO) in the Kerr metric, which depends on the black hole spin \citep[e.g.][]{gravitation}. For the hot corona, in the 2--10 keV band the spectral shape is a power law characterized by a photon index $\Gamma$. 
Finally, the main geometric parameter of the corona is the height above the event horizon of the black hole (while the size is computed by the code). A sketch of the lamppost configuration is given in Fig. \ref{fig:a}.
To summarize, the main input parameters are: the black hole mass and spin; the disc luminosity; the X-ray luminosity and spectral shape; the coronal height.
The details on the different parameters are described below.
\paragraph{\textit{Black hole mass and spin.}} For each source, the black hole mass is taken from the CAIXA catalog \cite[][]{caixa1}. The mass determines the Eddington luminosity $\ledd$ and the spectral shape of the emission from the Novikov-Thorne disc. 
Concerning the black hole spin $a$, we tested the two extreme cases $a=0$ (Schwarzschild) and $a=0.998$ \citep[maximally rotating;][]{thorne1974}. The inner disc radius is equal to 6 \rg\ for $a=0$ and to 1.24 \rg\ for $a=0.998$.
\paragraph{\textit{Disc luminosity.}}
The disc emission is modeled through the \kyn\ code \cite[]{kynbb}, which describes a blackbody-like spectrum from an accretion disc around a black hole, assuming a  Novikov-Thorne temperature profile. 
The spectrum is determined by two parameters, namely the black hole mass and the accretion rate or the luminosity in a given energy band.
We used the UV luminosity in the 5--7 eV range, roughly covering the bandpass of the UVM2 and UVW2 filters of the \xmm/OM. 
For such a small energy range, the luminosity is not strongly model-dependent. Therefore, for each source we estimated the 5--7 eV luminosity from the best fits of \cite{pics}, who assumed a multicolor disc blackbody. 
We note that \cite{pics} subtracted the main non-nuclear contributions, namely those from the host galaxy and the broad-line region; in any case, these components peak in the optical band and do not strongly contribute to the UV emission.
For the disc, we also assumed an inclination angle of 30 deg to the symmetry axis. However, this parameter does not strongly affect the results (see \ded).
\paragraph{\textit{X-ray spectrum and luminosity.}}
The X-ray emission from the corona is modeled through the \nthcomp\ code, whose relevant input parameters are:
\begin{itemize}
	\item the observed photon index;
	\item the electron temperature, fixed at 100 keV. The temperature sets the high energy cut-off, which affects the total X-ray flux especially for low photon indices;
	\item the normalization, given by the observed 2--10 keV luminosity;
	\item the temperature of the seed photons, estimated as $E\subrm{peak}/2.82$ where $E\subrm{peak}$ is the peak energy of the disc spectrum computed at the corona. This parameter sets the low energy cut-off, which affects the total X-ray flux especially for high photon indices.
\end{itemize}

We used the results of \cite{pics}, who fitted the \xmm/pn data with \nthcomp, also assuming a coronal temperature of 100 keV. 
The optical depths computed by the code are in the range 0.5--2, consistent with values commonly measured 
applying Comptonization models in spherical geometry (\citealt{tortosa2018}; see also \citealt{4388_2110}).

\paragraph{\textit{Coronal height.}} 
The center of the corona is at a height $H$ above the disc. We performed calculations assuming four different values, namely $H=2.5, 5, 10$ and 30 in units of gravitational radii.

\subsection{Results}\label{sec:results}
\begin{figure*}
   	\centering
   	\includegraphics[width=\hsize]{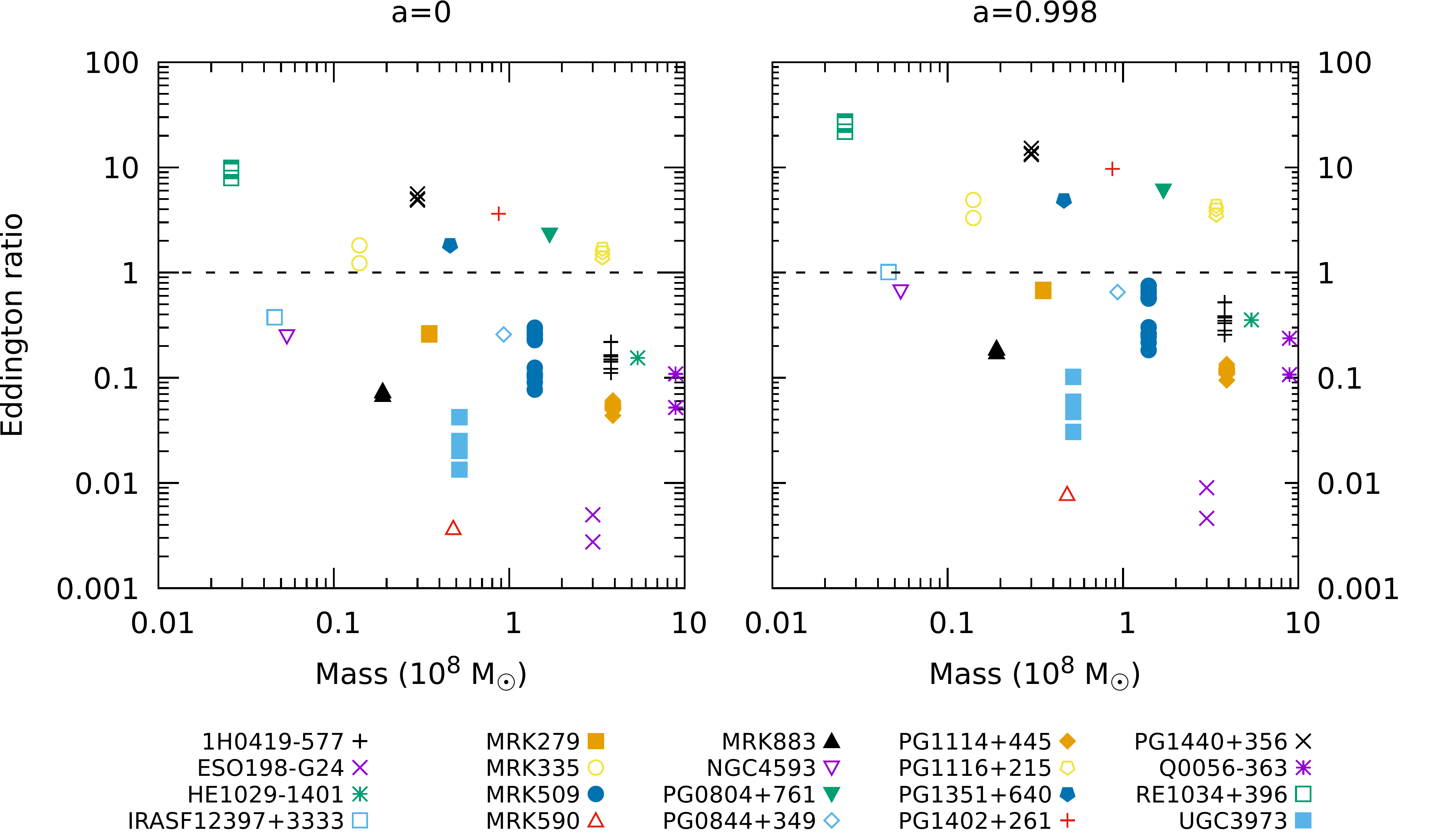}
   	\caption{Eddington ratio as calculated by the \ded\ code, assuming spin $a=0$ (left panel) and $a=0.998$ (right panel), plotted versus the black hole mass in units of \tento{8} solar masses.
   	}
   	\label{edd}
   \end{figure*}

\begin{figure*}
   	\centering
   	\includegraphics[width=\hsize]{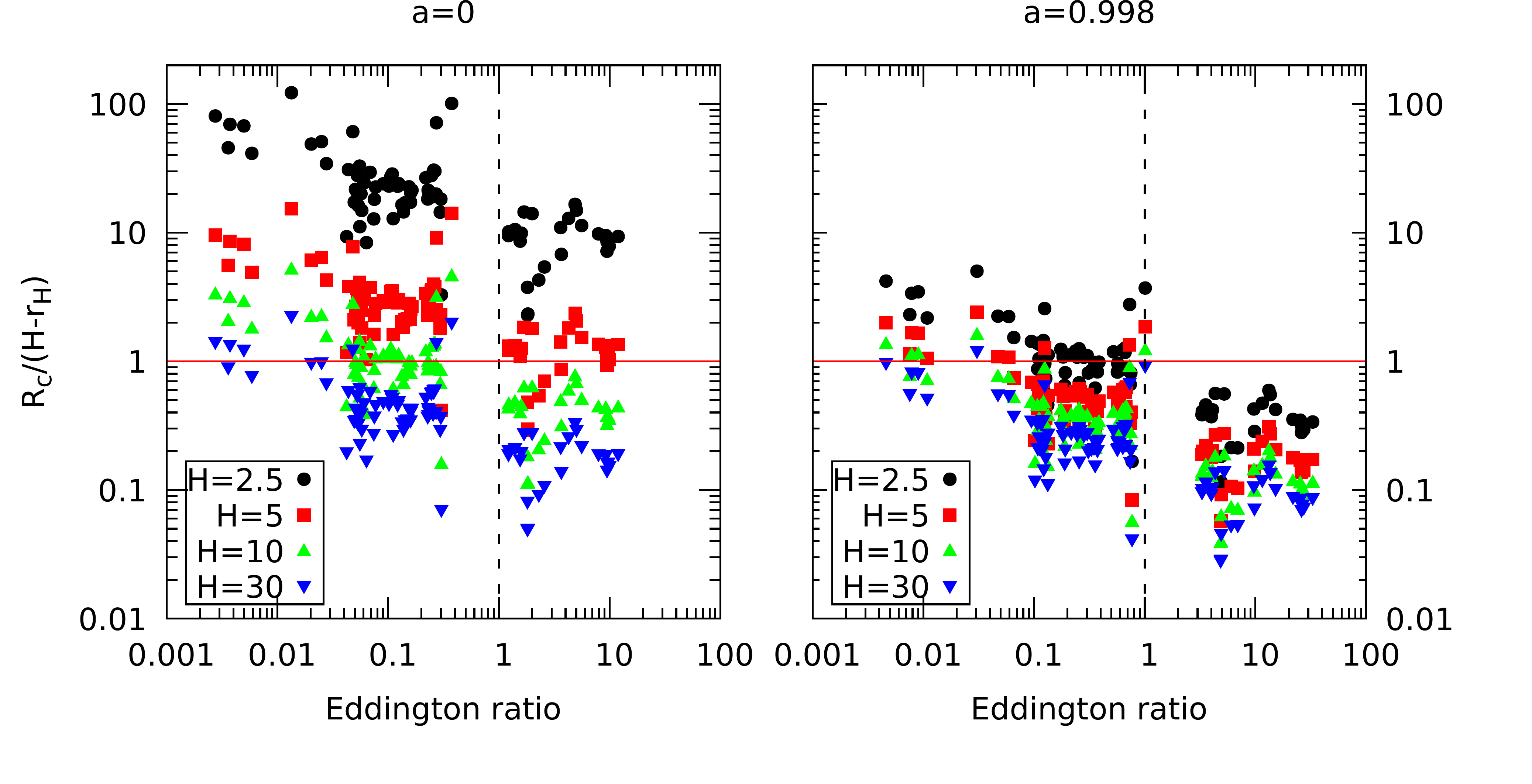}
   	\caption{Ratio of the coronal radius to the height above the event horizon plotted against Eddington ratio.
   	}
   	\label{fig:size}
   \end{figure*}

\begin{figure*}
      	\centering
      	\includegraphics[width=\hsize]{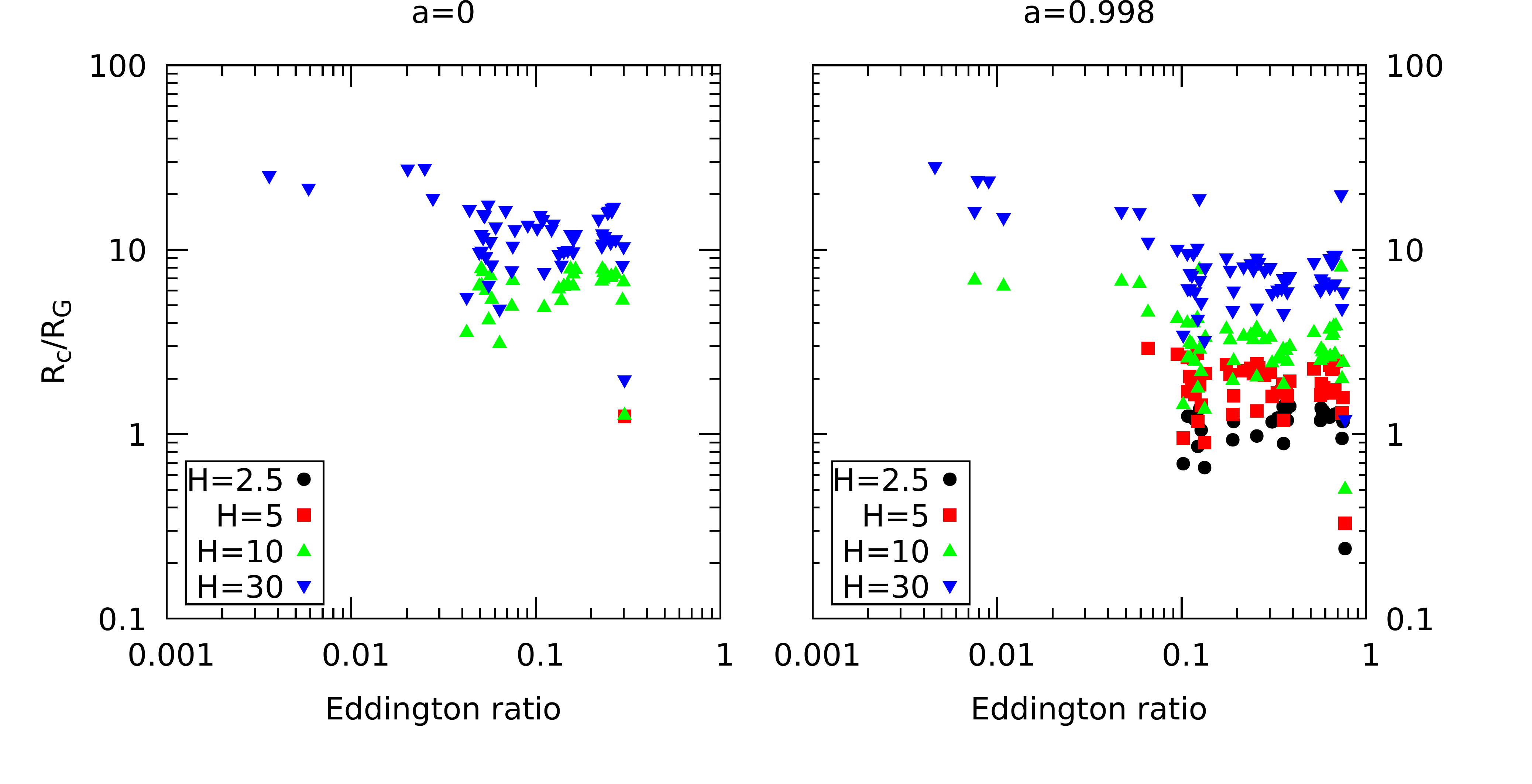}
      	\caption{Coronal radius $R_c$, in units of gravitational radii, versus Eddington ratio. We plot only those observations in which the corona can fit within the given height.
      	}
      	\label{fig:radius}
      \end{figure*}

\begin{figure*}
   	\centering
   	\includegraphics[width=\hsize]{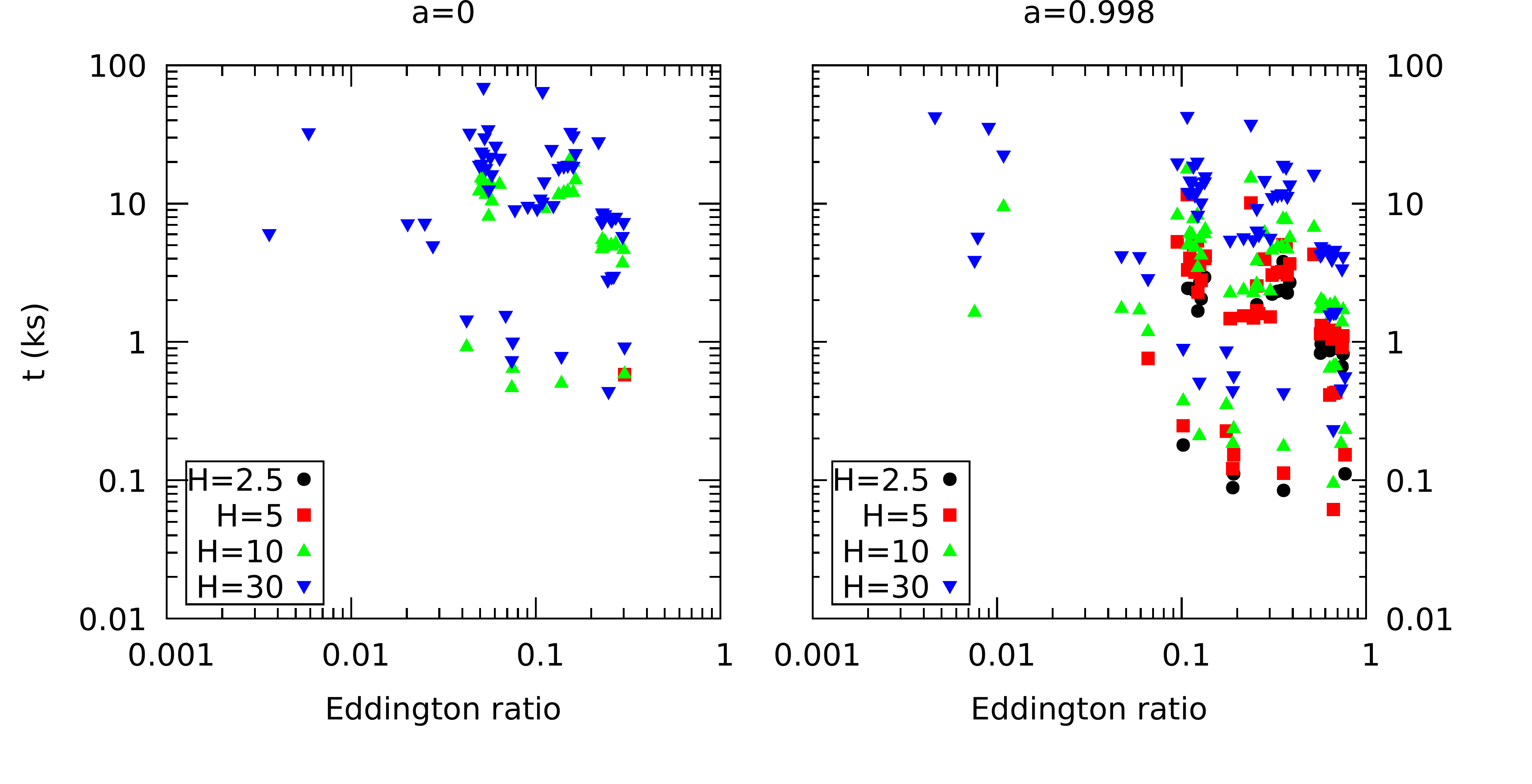}
   	\caption{Light-crossing time $t$ of the corona, in units of ks, versus Eddington ratio.
   	}
   	\label{fig:time}
   \end{figure*}

\paragraph{\textit{The Eddington ratio.}}
First of all, we checked the consistency between the output physical parameters calculated by the code and the assumed model.
In particular, the Eddington ratio $\eedd=L\subrm{bol}/\ledd$ (where $L\subrm{bol}$ is the bolometric luminosity of the disc) should not exceed unity, as this could conflict with the hypothesis of a standard Novikov-Thorne disc (see Sect. \ref{sec:conclusions}). 
The bolometric luminosity is the total luminosity of the \kyn\ component, integrated over the disc radius.  
In Fig. \ref{edd} we plot $\eedd$ versus the black hole mass. For a given mass, a larger value of the inner disc radius yields a smaller bolometric luminosity, because the disc has a smaller area while the observed UV luminosity is the same. As a result, the Eddington ratio for spin $a=0$ is smaller by a factor of 2-3 than for $a=0.998$. However, in both cases we obtain $\eedd \leq1$ for 13 sources out of 20. The highest Eddington ratio is found in the narrow-line Seyfert 1 RE~1034+396, with the caveat that the black hole mass is quite uncertain in this source \citep{czerny2016}. 
\paragraph{\textit{The corona size.}}
In Fig. \ref{fig:size} we show the ratio of the computed coronal radius to
the height above the horizon. For spin $a=0$, this ratio is greater than 1 for heights $H=2.5$ and 5 \rg, indicating that in these cases the corona cannot intercept enough seed photons to produce the observed X-ray flux. On the other hand, the corona can almost always fit within 30 \rg. For spin $a=0.998$, instead, the corona can almost always fit within 5 \rg, and even within 2.5 \rg\ in a limited number of sources. One major caveat is that the optical/UV emission of super-Eddington sources might be not well described by a Novikov-Thorne disc, as mentioned above.
The coronal radii of those sources with $\eedd \leq 1$ that are consistent with the lamppost geometry are plotted in Fig. \ref{fig:radius}. The corresponding light-crossing time $t$ in units of ks is plotted in Fig. \ref{fig:time}. 

\paragraph{\textit{The photon index.}}
Physical Comptonization models, that take into account the energetic coupling between corona and disc, predict a relationship between the X-ray spectral index and the Compton amplification factor $A$, namely the ratio between the total luminosity of the corona and the soft luminosity from the disc that enters the corona \cite[.e.g][]{haardt&maraschi1991}. 
In general, the Compton amplification depends crucially on geometry \cite[see also][]{pop2013mrk509,pics}. In the lamppost configuration, there is an inverse relation between the coronal radius and $A$: for a given luminosity, a bigger corona will intercept more disc photons, meaning that the Compton amplification factor will decrease. This is the basis for a self-consistency check \textit{a posteriori} of our results. Numerical simulations \cite[]{belo_discs,belo1999,malzac2001} show that the relationship between the photon index $\Gamma$ and the Compton amplification factor is fitted by a simple function:
\begin{equation}\label{eq:gamma}
\Gamma(A) = C (A-1)^{-\delta}
\end{equation}  
with $C\simeq 2$ and $\delta \simeq 0.1$. 
This relationship is obtained assuming a radiatively coupled disc-corona system, based on the balance between heating, via magnetic dissipation, and Compton cooling \cite[see also][]{haardt&maraschi1993}. 

\begin{figure*}
	\centering
	\includegraphics[width=\hsize]{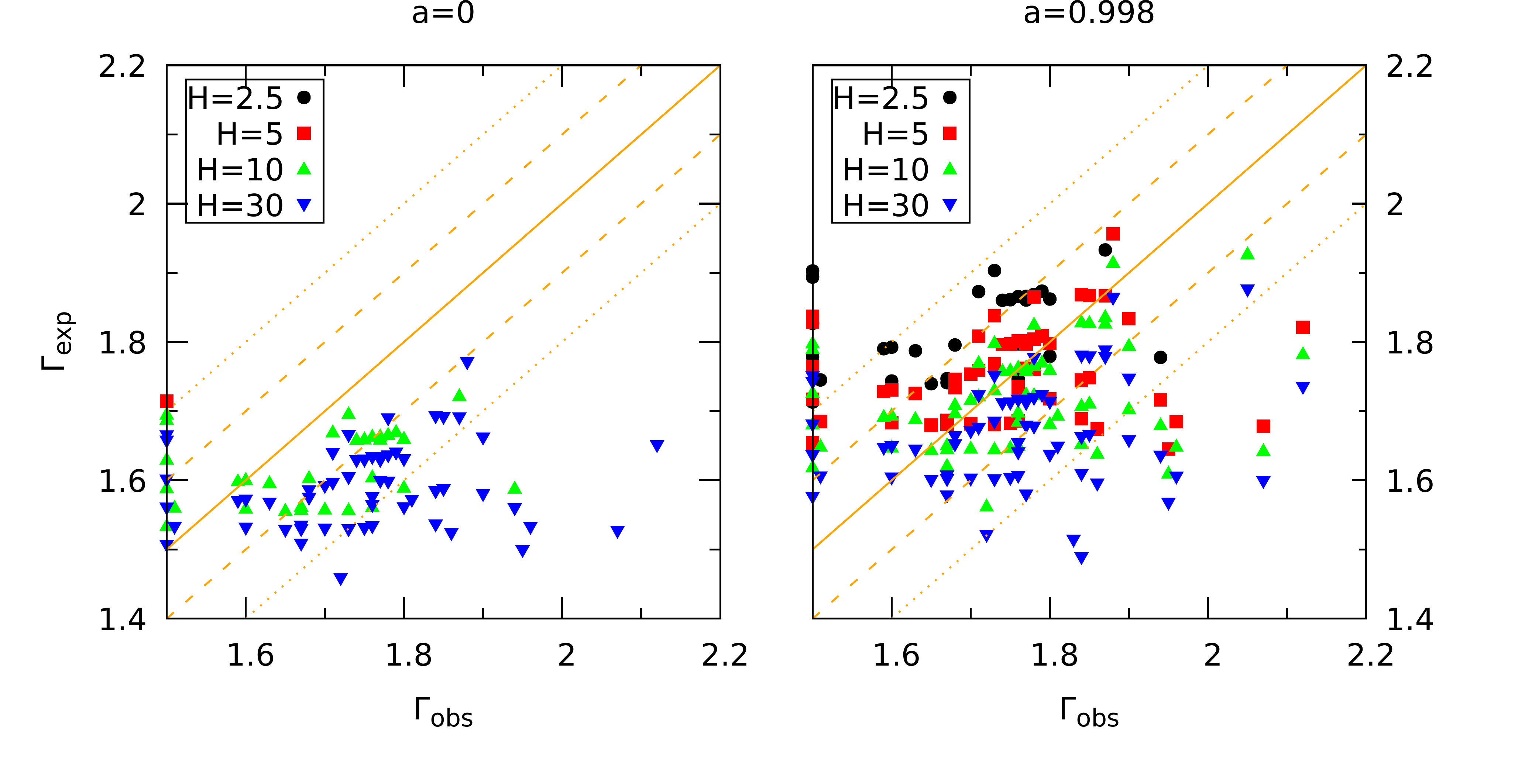}
	\caption{Expected photon index \cite[from eq. \ref{eq:gamma}, see][]{belo1999} versus observed photon index. The orange solid line represents the identity $\Gamma\subrm{exp} = \Gamma\subrm{obs}$, while the dashed and dotted lines correspond to differences $\Delta \Gamma = \pm 0.1, \pm 0.2$ respectively.
	}
	\label{fig:gamma}
\end{figure*}

The Compton amplification is calculated by the \ded\ code, from the flux of the disc photons intercepted by the corona.
We calculated the expected photon index (labeled $\Gamma\subrm{exp}$ in the following) from eq. (\ref{eq:gamma}) assuming $C=2.33$ and $\delta=0.1$ \cite[]{belo_discs,belo1999}\footnote{\cite{belo_discs,belo1999} performed calculations of dynamic coronae 
	applying both a simple analytical model and the numerical code of \cite{coppi1992} 
	in the case of a spherical blob. We obtain very similar results using the parameters of \cite{malzac2001}, who performed more detailed Monte Carlo simulations assuming a cylindrical geometry for the corona.}.
The expected photon index can be compared with the observed one ($\Gamma\subrm{obs}$). Since $A$ is computed within the code assuming $\Gamma\subrm{obs}$, we would expect $\Gamma\subrm{exp} \simeq \Gamma\subrm{obs}$, with the caveat that eq. (\ref{eq:gamma}) does not take into account general relativistic effects.
We plot in Fig. \ref{fig:gamma} $\Gamma\subrm{exp}$ versus $\Gamma\subrm{obs}$. 
For $a=0$, we mostly have $\Gamma\subrm{obs} > \Gamma\subrm{exp}$. On the other hand, for $a=0.998$ and height of 2.5 \rg\ we mostly have $\Gamma\subrm{obs} < \Gamma\subrm{exp}$. The best agreement is found in the case $a=0.998$ and intermediate heights (5 or 10 \rg).
\paragraph{} 
To summarize, we consider the lamppost corona on top of a standard disc to be consistent with the observations when: a) the Eddington ratio is $\leq 1$, b) the corona fits within at least one of the tested heights, and c) the difference $|\Gamma\subrm{exp}-\Gamma\subrm{obs}|$ is not too large. 
A reasonable condition is $|\Gamma\subrm{exp}-\Gamma\subrm{obs}| < 0.2$, keeping in mind that $\Gamma\subrm{exp}$ is the result of a simple analytical approximation and that the uncertainties on $\Gamma\subrm{obs}$ are in the range 0.01--0.20 \cite[]{pics}. 
We report in Table \ref{tab:constr} the sources that satisfy these constraints for all of their observations considered here. Conditions a) and b) are satisfied by eight sources for $a=0$ and eleven sources for $a=0.998$. Including condition c) leaves us with only two sources  for $a=0$ (Mrk~279 and Mrk~509) and five sources for $a=0.998$ (Mrk~279, Mrk~509, 1H~0419-577, HE~1029-1401 and NGC~4593).

      \begin{figure*}
      	\centering
      	\includegraphics[width=\hsize]{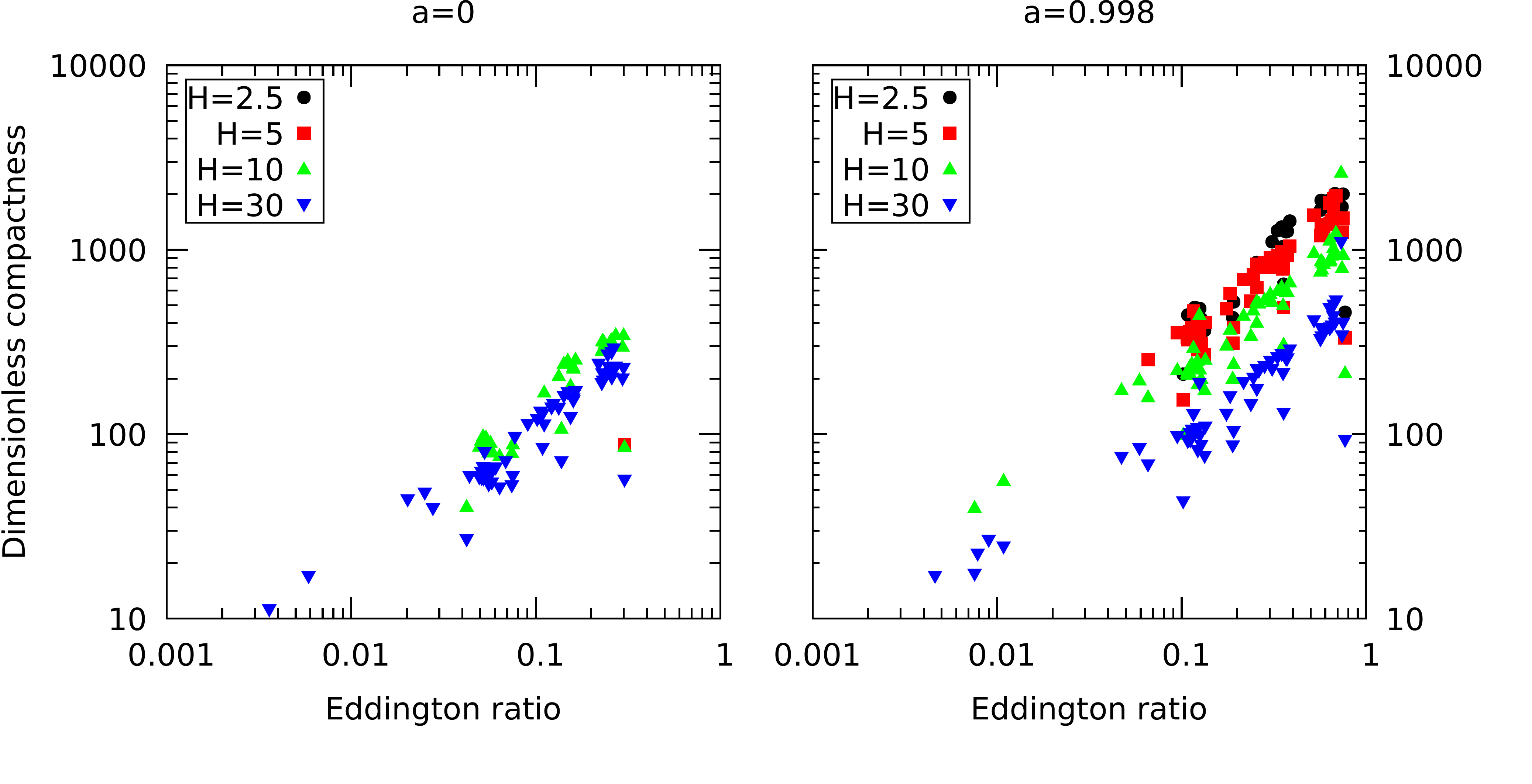}
      	\caption{Compactness $\ell$ versus Eddington ratio. 
      	}
      	\label{fig:l}
      \end{figure*}

\begin{table}
	\begin{center}
		\caption{Sources satisfying different constraints for all of their observations considered here (see Table \ref{tab:sample}): a) $\eedd \leq 1$, b) corona fits within at least one height, c) $|\Gamma\subrm{exp} - \Gamma\subrm{obs}| <0.2$. \label{tab:constr}}
		\begin{tabular}{lcc|cc}
			\hline \hline
			&\multicolumn{2}{c}{$a=0$}&\multicolumn{2}{c}{$a=0.998$}\\
			\hline
			a) &b) & c)&b) & c)\\
MRK 279	&	\ding{51}	&	\ding{51}	&	\ding{51}	&	\ding{51}	\\
MRK 509	&	\ding{51}	&	\ding{51}	&	\ding{51}	&	\ding{51}	\\
1H 0419-577 	&	\ding{51}	&	\ding{55}	&	\ding{51}	&	\ding{51}	\\
HE 1029-1401	&	\ding{51}	&	\ding{55}	&	\ding{51}	&	\ding{51}	\\
NGC 4593	&	\ding{55}	&		&	\ding{51}	&	\ding{51}	\\
MRK 883	&	\ding{51}	&	\ding{55}	&	\ding{51}	&	\ding{55}	\\
PG 0844+349	&	\ding{51}	&	\ding{55}	&	\ding{51}	&	\ding{55}	\\
PG 1114+445	&	\ding{51}	&	\ding{55}	&	\ding{51}	&	\ding{55}	\\
Q0056-363	&	\ding{51}	&	\ding{55}	&	\ding{51}	&	\ding{55}	\\
MRK 590	&	\ding{55}	&		&	\ding{51}	&	\ding{55}	\\
ESO 198-G24	&	\ding{55}	&		&	\ding{51}	&	\ding{55}	\\
IRASF 12397+3333	&	\ding{55}	&		&	\ding{55}	&		\\
UGC 3973	&	\ding{55}	&		&	\ding{55}	&		\\
			\hline
			tot. 13 &8&2&11&5 \\
			\hline
		\end{tabular}
	\end{center}
\end{table}

\section{Discussion and conclusions}\label{sec:conclusions}
The estimates of the \ded\ code indicate that, at least for the sub-Eddington sources, the lamppost Comptonizing corona is generally a viable scenario. However, such a corona can only exist beyond a certain height above the event horizon, depending on inner disc radius. If the disc is extended down to the ISCO of a maximally spinning black hole,
the corona can be located down to a height $H=2.5$ \rg, having a radius smaller than $H-r\subrm{H} =1.44$ \rg, at least for sources with $\eedd \gtrsim 0.1$ (see Fig. \ref{fig:size}).  
A coronal height of 5 \rg\ is a physically consistent value in almost all sources analyzed here. On the other hand, for a non-spinning black hole, the corona cannot fit within a height of 5 \rg. For $\eedd \gtrsim 0.04$, the corona can be at $H=10$ \rg, having a radius smaller than $H-r\subrm{H} =8$ \rg. 

The observed spectral shape can also provide some constraints on the coronal geometry, comparing the observed photon index with that predicted from the Compton amplification.
The best agreement between observed and expected photon indices is found for maximal spin and heights of 5 \rg\ or larger. 
This solution is thus favored, while assuming a low spin predicts a spectrum mostly flatter than observed for the sources considered here. On the other hand, for a high spin and a height of 2.5 \rg\ the predicted spectrum is mostly steeper than observed.
We note that the number of sources with a spectral shape roughly consistent with the expectation, for all of their observations, is not high: if $a=0.998$, these sources amount to 5 out of the 11 for which the corona can geometrically fit. This could indicate that a spherical corona on the disc axis might not always represent a good model. Future X-ray polarimetric observations will be crucial to constrain the geometry and distinguish between the possible configurations of the corona \cite[]{moca1,marinucci2019}.

From the coronal radii, we calculated the dimensionless compactness parameter $\ell \equiv L \sigma_\textrm{T} / R_c m\subrm{e} c^3$, where $L$ is the luminosity of the corona (see the plot in Fig. \ref{fig:l}). Following \cite{fabian2015coronae}, we computed the luminosity in the 0.1--200 keV band
of the \nthcomp\ model (very close to a power law with a sharp cut-off at 100 keV).
The corona is always consistent with being radiatively compact, in the sense that $\ell$ is always $>10$. We mostly find estimates of $\ell$ between $\sim 100$ and $\sim 1000$, consistent with the observational constraints that are generally found with \nus\ \citep{fabian2015coronae,fabian2017}. 
For a given compactness, there exist a maximum temperature below which pair equilibrium is possible; above this temperature, pair production becomes a runaway process. Therefore, especially in the case of large compactness, the observation of the X-ray source strongly constraints the temperature, because even a few photons above 511 keV are sufficient to produce runaway pair production. We note that our assumption of a 100 keV temperature 
is consistent with pair equilibrium, because it would place the sources below the pair runaway line for a spherical corona \citep{fabian2015coronae,fabian2017}.

The main limitation of the model used in this work is likely the assumption of a standard, geometrically thin accretion disc. Indeed, super-Eddington sources are thought to be powered by advection-dominated slim/thick discs, rather than by standard thin discs \cite[e.g.][]{pw1980,slim,mineshige2000,seambh1,castello2016}. On the other hand, an Eddington ratio much lower than unity might suggest the presence of a radiatively inefficient accretion flow \cite[e.g.][]{adaf}. However, at least for the sources having $\eedd$ in the range $0.01-1$, the standard disc assumption is realistic \cite[e.g.][]{czerny2018}.
We note that \cite{pics} discuss a physical scenario in which the disc is sandwiched by a warm Comptonizing corona, with a temperature in the range 0.1--1 keV, to account for the soft X-ray excess seen in many unobscured Seyfert galaxies \cite[see also][]{middei_4593,he1143}. However, the photon rate from the warm corona, hence the distribution of seed photons seen by the hot corona, is always dominated by optical--UV photons. Therefore, assuming a standard disc as the illumination source for the hot corona would be a fair approximation also in this case. 

We also note that the model does not take into account the disc-corona energetic coupling, in particular the fact that a fraction of the accretion power has to be channeled into the corona to provide its heating, thus reducing the disc flux by the same amount. However, this should not strongly alter our results, because the X-ray luminosity is a generally a small fraction ($10\%$ or less) of the bolometric luminosity.
The coronal radiation reflected by the accretion disc is also not currently computed by the code, and will be implemented in a future work. The reflection component can, in principle, carry significant information on the geometry of the corona. For example, the profile of the ubiquitous \fek\ emission line at $\sim 6.4$ keV can be broadened and skewed \cite[e.g.][]{nandra2007pexmon}, which is often interpreted as a general relativistic effect of reflection off the inner disc (e.g. \citealt{tanaka1995}, but see also \citealt{miller2008}).  
Combining spectral and timing analysis of the X-ray continuum and reflection component can yield constraints on the height of the corona, the inner disc radius and their temporal evolution: for instance, \cite{kara2019} report a shrinking of the corona in a stellar-mass black hole \cite[but see also][]{mahmoud2019,kajava2019}, while \cite{caballero2020} report coronal height variations in a narrow-line Seyfert 1. 
Further studies will be needed to understand the dynamical properties of the corona.

Another source of uncertainty is the black hole mass, because the luminosity and the spectral shape of the disc emission, as well as the Eddington ratio, depend crucially on this parameter.  Around $50\%$ of the objects in our sample have reverberation-based measurements of the black hole mass, while the other measurements are based on the width of the H $\beta$ emission line \cite[]{caixa1,caixa3}. The former method yields mass uncertainties of a factor of $\lesssim 3$ \cite[]{peterson2004}, but the latter method yields larger uncertainties \cite[]{caixa3}. This can potentially affect the results. For example, \cite{done&jin_1h} discussed in detail the case of the narrow-line Seyfert 1 1H~0707-495. For this source, the spectral-timing models based on relativistic reflection in the lamppost geometry require a high black hole spin ($a = 0.998$) and a mass of \expo{2}{6} $\msun$ \cite[]{fabian2012_1h}. These parameters imply a highly super-Eddington regime. However, \cite{done&jin_1h} found also a sub-Eddington solution assuming low spin ($a = 0$) and a mass of \expo{1}{7} $\msun$. In general, we can expect smaller accretion rates if the black hole masses are larger than currently estimated, so the corona size could in turn be underestimated in these cases. Future extensions of our results will be possible as further, robust measurements of black hole masses become available.

\begin{acknowledgements}
We thank the anonymous referee for comments that improved the manuscript.
We also thank S. Bianchi, A. De Rosa and A. Marinucci for useful discussions.
MD and WZ acknowledge the financial support provided by the Czech Science Foundation grant 17-02430S and institutional support by the project RVO:67985815.
GM  acknowledges  financial support from the
Italian Space Agency under grant n. 2017-14-H.O.	
POP acknowledges financial support from CNES and the CNRS/PNHE.
CD acknowledges the Science and Technology Facilities Council (STFC)
through grant ST/P000541/1.
\end{acknowledgements}

\bibliographystyle{aa}
\bibliography{mybib.bib}
\end{document}